\documentstyle[multicol,aps,prl,graphicx]{revtex}

\begin{document}
\draft
\title{Dynamical mean-field theory of electron-phonon interaction in correlated
electron materials: general results and application to doped Mott insulators}
\author{Andreas Deppeler$^1$ and Andrew J. Millis$^2$}
\address{$^1$Center for Materials Theory, Department of Physics and Astronomy, Rutgers University, Piscataway, New Jersey 08854 \\
$^2$Department of Physics, Columbia University, 538 W 120th St, New York, New York 10027}
\date{\today}
\maketitle

\begin{abstract}
The dynamical mean-field method is used to formulate a computationally
tractable theory of electron-phonon interactions in systems with arbitrary 
{\it local} electron-electron interactions in the
physically relevant adiabatic limit of phonon frequency small compared to
electron bandwidth or interaction scale. As
applications, the phonon contribution to the effective mass of a carrier in
a lightly doped Mott insulator is determined and the phase separation
boundary is discussed.
\end{abstract}


\begin{multicols}{2}

`Correlated electron' materials such as doped Mott insulators \cite{Imada98},
high-temperature superconductors \cite{Orenstein00}, heavy fermions \cite
{Wilkins86}, and `half-metallic' oxides \cite{Salamon01} are central to
present-day condensed matter physics because they exhibit many properties
that seem incompatible with predictions of the standard paradigm
[local density approximation (LDA) band theory plus Boltzmann transport] of
metal physics. A crucial role is played 
by electron-electron interaction effects beyond the scope of the LDA, and
theoretical attention has focused on these effects
neglecting other physics. However, the electron-phonon
interaction is present in all real materials, and it is therefore important
to understand its effects in correlated systems. The relevance of
electron-lattice effects in the CMR manganites is by now well established 
\cite{Salamon01,Millis96b}. Lanzara and co-workers presented
photoemission data which, they argue, indicate
that electron-phonon effects play an important role in 
high-temperature superconductivity \cite{Lanzara01}. On
the other hand, the apparent absence of electron-phonon
effects in the resistivity of high-$T_c$ materials
is a long-standing mystery \cite{Gurvitch87}.

There is to date no systematic theoretical extension to correlated materials of the
successful Migdal-Eliashberg (ME) theory \cite{Migdal58},
which describes electron-phonon interactions in {\it weakly correlated} materials. 
Electron-phonon interactions in one-dimensional systems have been
studied by renormalization group methods \cite{Bourbonnais89},
and recent improvements in numerical techniques have for example
allowed the spin-phonon dynamics of insulating quasi-one-dimensional Peierls
systems to be determined in considerable detail \cite{Trebst01}. Concerning
higher-dimensional, metallic systems, Kim
and co-workers \cite{Kim89} argued that near a Mott transition
a decrease in electron-phonon coupling was compensated by an
increase in carrier mass, leading to electron-phonon effects
unrenormalized by proximity to the Mott transition.
Our results, to be presented below, disagree with this conclusion. 
Several authors have used direct numerical simulation
[e.g., quantum Monte Carlo (QMC) techniques] of models of electrons
interacting both with each other and with lattice vibrations \cite
{Pao98,Berger95,Freericks93}. However, the present limitations of memory size
and algorithms have restricted these works mainly to the calculation of
static properties, especially phase boundaries and transition temperatures
and to the `antiadiabatic' limit of phonon frequency comparable to electronic energy
scale.

The development of the
dynamical mean-field (DMF) method \cite{Georges96} has opened an
important avenue for progress, by showing that if 
(as occurs for the electron-phonon interaction) the
momentum dependence of the electron self-energy is negligible then a good
approximation to the correlation physics can be obtained from the solution
of a numerically tractable quantum impurity problem plus a self-consistency
condition. Unfortunately the straightforward inclusion of the
electron-phonon coupling in the DMF formalism is difficult because
the mismatch between the typical phonon frequency scale $\omega
_{0}\lesssim 0.1$ eV and electron energy scale $t\gtrsim 1$ eV renders
conventional numerical approaches to the impurity problem prohibitively
expensive, except in the `antiadiabatic' limit \cite{Freericks93} 
of relevance to rather few materials. In this paper we present a
practical implementation of an
adiabatic expansion of the DMF formalism and show how it
may be used to determine the phonon contribution to electronic properties of
correlated systems. The work reported here builds on previous papers, which
introduced the adiabatic expansion \cite{Deppeler02a} and applied it to models involving only
electron-lattice interactions \cite{Blawid01,Deppeler02b}.
As an application we answer the long-standing question of
the electron-phonon contribution to the carrier self-energy in a lightly
doped Mott insulator. Our methods may easily be extended, 
e.g., to heavy fermion materials and may be combined with recent
extensions of the DMF method \cite{Anisimov97,Held01}.

We consider the single-site DMF approximation
to a general Hamiltonian of the form $H=H_{{\rm band}}+H_{{\rm el-el}}+H_{%
{\rm ph}}+H_{{\rm el-ph}}$, where $H_{{\rm band}}$ and $H_{{\rm el-el}}$
describe electrons moving in some band structure and interacting via some
local interaction. We model the phonons as dispersionless quantum oscillators 
with instantaneous displacement $q^a$, mass $M_a$, and spring
constant $K_a$: 
\begin{equation}
H_{{\rm ph}}=\frac{1}{2}\sum_{i\,a}[M_{a}\left( \partial _{\tau
}q^{a}\right) _{i}^{2}+K_{a}q_{i}^{a2}].  \label{Hph}
\end{equation}
Dispersion may be approximately included by treating $M,K$ as Brillouin-zone averages of
phonon dispersions or via the more sophisticated techniques of Refs. \cite
{Motome00} and \cite{Jarrell01}. Anharmonic
terms such as those considered in Ref. \cite{Freericks00} can be
easily added and will be seen to be generated.

The electron-phonon coupling is again taken  to be
local. It is cumbersome to write  in  general. We illustrate
the issues via  the physically relevant example of the
pseudocubic manganese perovskites. Here the relevant electrons are $e_g$%
-symmetry Mn-O hybrid states. The generalized phonon field $q_{a}$
encompasses (i) a `breathing' mode (symmetric distortion of the Mn-${\rm O}_{6}$
octahedron) coupling via a constant $g_{{\rm B}}$ to the total on-site
charge density and (ii) a one-parameter family of Jahn-Teller
modes (even-parity,
volume-preserving distortions of the Mn-${\rm O}_{6}$ octahedron) coupling via a
matrix element $g_{{\rm JT}}$ to appropriate differences of
occupancy between different orbitals (for a more detailed discussion of the
couplings and notation see, e.g, Ref. \cite{Millis96b}). These two distortions will
be labeled by a scalar coordinate $x$ and a two-component vector $\vec{Q}%
=(Q_{x},0,Q_{z})$, respectively. The coupling term reads (repeated
indices are summed)
\begin{equation}
H_{{\rm el-ph}}=\sum_{i}[ g_{{\rm B}} x_{i}(n_{i}-n)+g_{{\rm JT}}%
\vec{Q}_{i}\cdot c_{ia\beta }^{\dagger }\vec{\tau}^{ab}c_{ib\beta }],
\label{epi}
\end{equation}
where $\vec{\tau}=(\tau _{x},\tau _{y},\tau _{z})$ are Pauli
matrices acting in orbital space, 
$n_{i}=\sum_{a\alpha }c_{ia\alpha }^{\dagger }c_{ia\alpha }$ is
the local electron density, and $\langle x_{i}\rangle =0$ is defined to be
the equilibrium phonon state for a uniform distribution of electrons.

Within the single-site DMF approximation \cite{Georges96} the properties 
of $H$ may be obtained from the solution of an impurity model specified by the
action $S[c,\bar{c},q,a] = S_{0}[q] + S_{1}[c,\bar{c},q,a]$, with 
$S_{0}[q] = 1/(2T)\sum_{k}q_{k}^{a}\left( K_{a} + M_{a}\omega
_{k}^{2}\right) q_{-k}^{a}$ and $S_1[c,\bar{c},q,a] =
S_{{\rm el-el}}[c,\bar{c}] + S_{{\rm el-ph}}[c,\bar{c},q] - \sum_{\alpha ,\,n}\bar{c}_{\alpha n}c_{\alpha n}a_{n}$. Here the terms
$S_{\rm el-el}$ and $S_{\rm el-ph}$ are obtained in the usual way from the interaction terms
listed above, and $S$ is a functional of a mean-field function $a$, which may
depend on spin and orbital indices and expresses the effect of the rest of
the lattice upon the single site. It is fixed by equating
the impurity Green function $G_{{\rm imp}}[a]_{n} = \delta\ln Z[a]/\delta a_n$
to the local Green function $G_{{\rm loc}}[a]_{n} = \int d^3p/(2\pi)^3\,
(i\omega _{n}+\mu -\Sigma[a]_{n}-H_{\rm band})^{-1}$
with $\Sigma[a]_{n} = a_n - G_{\rm imp}[a]_n^{-1}$. The fields $c,\bar{c}$ and $q$ 
represent local electronic and phonon degress of freedom and have been
Fourier transformed according to $q(\tau )=\sum_{k}\exp (-i\omega _{k}\tau
)q_{k}$ etc.  We will index bosonic Matsubara frequencies $\omega
_{k}=2k\pi T$ by integers $k$ and fermionic Matsubara frequencies $\omega
_{n}=(2n+1)\pi T$ by integers $n$, i.e., $q_{k}\equiv q(i\omega _{k})$ and $%
a_{\alpha n}^{-1}\equiv a_{\alpha }^{-1}(i\omega _{n})$. 

To analyze the effect of phonons on electronic physics \cite
{Deppeler02a,Blawid01} we integrate out the electron fields and
work with an effective action $S[q,a]=S_{0}[q]+S_{1}[q,a]$,  
which we expand about the values $\bar{q}$ that extremize $S$. 
We introduce an electronic bandwidth or interaction scale $t$ and define the parameters
\begin{equation}
\lambda =\frac{g^{2}}{Kt},\qquad \gamma =\frac{(K/M)^{1/2}}{t}=\frac{\omega
_{0}}{t}.\label{fpar}
\end{equation}
Crucial objects in the expansion are the vertices 
\begin{equation}
\Gamma_{N}[a]_{k_{1},\ldots ,k_{N}} = -\frac{1}{(N-1)!}\,\frac{\delta S_{1}[q,a]}{\delta \,q^{a_1}_{k_{1}}\cdots \delta \,q^{a_N}_{k_{N}}},  \label{gna}
\end{equation}
which are connected correlation functions of the electrons-only theory. The
adiabatic parameter $\gamma $ controls the expansion because each
phonon loop involves a sum over frequencies of order $\omega_0 \sim \gamma t$ 
while the vertices $\Gamma$ vary with frequency on the scale $t$. If terms of order $\gamma ^{3/2}$ and
higher are neglected, the $\Gamma _{N>4}$ may be dropped and $\Gamma _{3,4}$ may be approximated by their static value computed using $a$ at $\gamma=0$. With suitably rescaled fields and frequencies (for details see Ref. \cite{Deppeler02a}) we may write
\begin{eqnarray}
S[q,a] &=&\frac{1}{2}\sum_{k}q_{k}D_{k}^{-1}q%
_{-k}  \nonumber \\
&-&\frac{\lambda ^{3/2}\gamma ^{1/2}T^{1/2}}{3}\Gamma%
_{3}[a]\sum_{k_{1},k_{2}}q_{k_{1}}q_{k_{2}}q%
_{-k_{1}-k_{2}}  \nonumber \\
&-&\frac{\lambda ^{2}\gamma T}{4}\Gamma_{4}[a]%
\sum_{k_{1},k_{2},k_{3}}q_{k_{1}}q_{k_{2}}q_{k_{3}}%
q_{-k_{1}-k_{2}-k_{3}},  \label{efalo}
\end{eqnarray}
where $D_{k}$ is the phonon propagator, equal for
scalar phonons to $1/(1+\omega_{k}^{2}/\omega_{0}^{2}-\lambda \Gamma _{2}[a]_{k,-k})$, 
renormalized by electron-electron
interaction effects contained in $\Gamma _{2}$. Note that 
${\cal O}(\gamma )$ corrections to $a$, as well as the leading 
frequency dependence, must be included in $\Gamma _{2}$.

\begin{figure}
\includegraphics[width=3.0in]{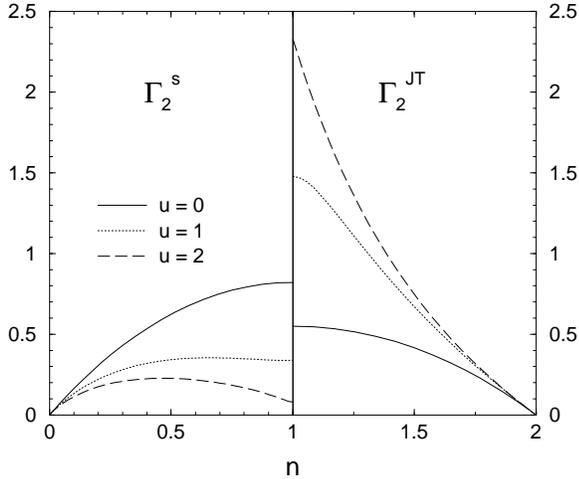}
\caption{Left panel: quadratic vertex $\Gamma_2$ for scalar phonons as a
function of electron density $n$, for various values of electron-electron
interaction $u := U/(2t)$ and $\protect\beta = 10$. Right panel:
$\Gamma_2$ for Jahn-Teller phonons in the paramagnetic 
(i.e., {\it spin-disordered\/}) phase of the 
Hubbard--double-exchange model $H_{\rm el-el}=Un(n-1)+J\vec{S} \cdot \vec{\sigma}$
with $J \rightarrow \infty$. Numerical method: QMC with $L = 32$ time slices.}
\label{G2JTn}
\end{figure}

\begin{figure}
\includegraphics[width=3.0in]{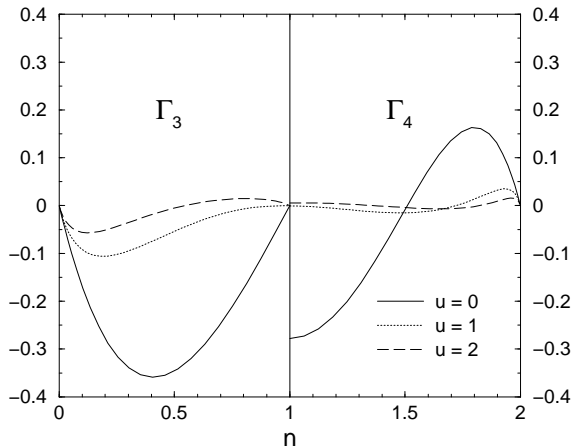}
\caption{Left panel: cubic vertex $\Gamma_3$ 
for single-band Hubbard model at $\protect\beta = 10$, with $u$ and 
electron density $n$ as shown. Right panel: quartic vertex $\Gamma_4$ for same parameters.
Numerical method: exact enumeration with $L = 16$.
Note: $\beta=10$ is higher than the critical endpoint of the Mott transition
occurring at $u \approx 1.5$.}
\label{G34n}
\end{figure}

We used the Hirsch-Fye \cite{Hirsch86} algorithms distributed
with Ref. \cite{Georges96} and routines to compute
the connected correlation functions for 
a single-band Hubbard model. Computations were performed
on a Sun workstation; the longest computation $\Gamma_2$ at
larger $U$ took $\sim 40$ minutes. Errors (estimated by
comparison to exact solution and scaling with system sizes)
were less than $2\%$ for parameters shown.
Fig. \ref{G2JTn} shows that near half filling and for scalar phonons
the Hubbard-$U$ strongly decreases $\Gamma_2$, thereby 
suppressing the polaronic instability occurring at 
$\lambda \Gamma_2=1$. Although the polaronic instability
is suppressed a phase separation instability is found for
$\lambda \gtrsim 1$ and $n$ near 1; details will be presented
elsewhere. For comparison and to demonstrate
the power of the method we also present 
in the right-hand panel results for a model
of `colossal magnetoresistance' manganites in which orbitally degenerate electrons
are coupled to Jahn-Teller (JT) phonons via the second term in Eq. 
(\ref{epi}) and also feel both a Hubbard-$U$ and a 
`double-exchange' interaction arising from a strong coupling
to core spins (i.e., the model defined in Ref. \cite{Millis96b} but with a Hubbard-$U$ added).
The $U$ is seen to enhance the effect of JT phonons.
Fig. \ref{G34n} shows that for the Hubbard model
$\Gamma_3$ and $\Gamma_4$ are even more rapidly suppressed
by $U$ so repulsive interactions effectively decouple
electrons from scalar phonons.

\begin{figure}
\includegraphics[width=3.0in]{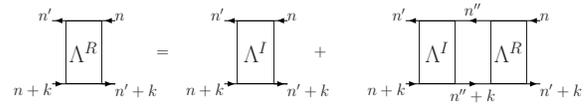}
\caption{Diagrammatic Dyson equation connecting the reducible and
irreducible particle-hole vertices $\Lambda^R$ and $\Lambda^I$,
respectively. }
\label{verir}
\end{figure}

We now turn to the electron self-energy $\Sigma ^{{\rm ph}}$ due to
interactions with phonons, assuming that the effective vertices
$\Gamma_{N}$ have been determined as discussed above. 
The calculation is simplified
and made more physically transparent if the action is viewed as a
functional of the full electron propagator $G$ rather than as a functional
of $a$. The key object is then the
`local Luttinger-Ward functional' $\phi[G]$ defined \cite{Abrikosov75} as
the sum of all vacuum-to-vacuum skeleton diagrams and related to  the self-energy via $\Sigma =\delta \phi[G]/\delta G$. Within DMF theory, $\phi $ may be found 
from  the local thermodynamic potential $\Omega_{\rm imp}=-T\ln Z$ (see Refs. \cite{Georges96} and \cite{Brandt91}) so may be computed by an adiabatic expansion. We find
 $\phi[G] = \phi^{\rm el-el}[G] + \phi^{\rm ph}[G] + {\cal O}(\gamma ^{3/2})$
with $\phi^{{\rm ph}}[G]=\frac{1}{2}\sum_{k}\ln D_{k}^{-1}$. Therefore
\begin{equation}
\Sigma _{n}^{{\rm ph}}=-\frac{\lambda }{2}\sum_{k}D_{k}\frac{\delta 
\Gamma_{2}[G]_{k,-k}}{\delta G_{n}}+{\cal O}(\gamma ^{2}).
\end{equation}
In general $\Gamma _{2}$ and thus $\delta \Gamma _{2}/\delta G$
vary on the scale of $t$ so $\Sigma^{\rm ph} \sim \gamma $, an unimportant
correction to the bare frequency or electron-electron contribution. 
However, if $\delta \Gamma/\delta G$ is singular at low frequencies,
as in Fermi liquids, then $\partial \Sigma /\partial \omega $ may be of
order unity. To investigate this we note
that $\Gamma_{2}[G]_{k,-k}=-2tT\sum_{n}G_{n}G_{n+k}\left(1+2tT\sum_{n^{\prime }}\Lambda
_{nn^{\prime }k}^{R} G_{n^{\prime }}G_{n^{\prime }+k}\right)$, where the particle-hole {\em %
reducible\/} vertex $\Lambda ^{R}$ is given in terms of the particle-hole 
{\em irreducible\/} vertex $\Lambda ^{I}$ via 
the Dyson equation shown in Fig. \ref{verir}.

The standard analysis implies that $\Lambda ^{I}$ is a smooth function of its arguments,
so the required singular behavior can only occur if we differentiate on one
of the explicit $G$ factors. Further, to ${\cal O}(\gamma)$ we may neglect
the frequency dependence of the vertices leading to 
\begin{equation}
\Sigma _{n}^{{\rm ph}}=\lambda tT\sum_{k}D_{k}G_{n+k}\Lambda^{2}+\Sigma _{n}^{{\rm ph\,reg}}  \label{sigme}
\end{equation}
with $\Sigma _{n}^{{\rm ph\,reg}}$ varying with $\omega_n$ on the scale of $t$ and
\begin{equation}
\Lambda = 1 + 2tT\sum_{n}\Lambda_{0n0}^{R}G_{n}^2 + {\cal O}(\gamma). \label{lkren}
\end{equation}
If the ground state is a Fermi liquid then $G(i\nu )=-i\pi \,{\rm sign}(\nu
)\rho (\mu )+G^{{\rm inc}}$, where $G^{{\rm inc}}$ is nonsingular at small
frequencies \cite{Rozenberg94}. The leading contribution in $\gamma $ to
$m^{\ast}/m$ is then $m^{\ast}/m|_{\rm ph} = 1 + \bar{\lambda}t\rho(\mu )\Lambda^{2} + {\cal O}(\gamma )$, where $\bar{\lambda} = \lambda/(1 - \lambda\Gamma_2)$.

\begin{figure}
\includegraphics[width=3.0in]{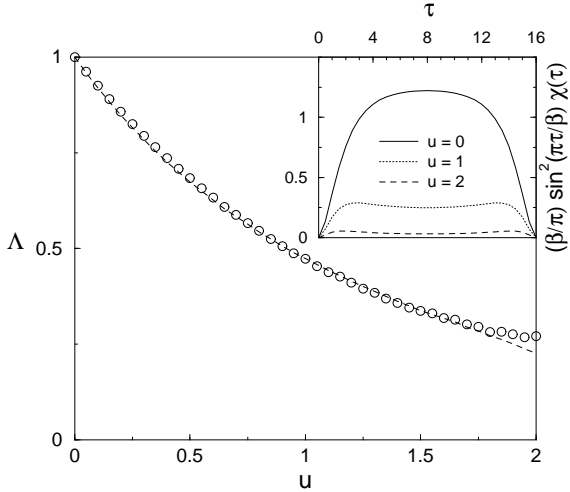}
\caption{Static particle-hole scattering renormalization factor as function
of $u := U/(2t)$. Numerical results for $\protect\beta = 16$ and $\protect%
\mu = 0$. Circles: data as described in text. Dashed line: analytic
weak-coupling approximation. Inset: imaginary time
dependence of $\chi$ showing accuracy of approximation.}
\label{lam0u}
\end{figure}

$\Lambda$ is a vertex function of the
underlying electron-electron theory, determined in general by
solving the vertex equation shown in Fig. \ref{verir}. However, to the order to
which we work it is simpler to note that in a Fermi liquid the leading low-frequency behavior of the density-density correlation function is $\chi (\omega) = \chi(0) + iA |\omega|$ with $A=\Lambda
^{2}\partial \chi^0/\partial \omega$ and $\chi^0_k = -2tT\sum_{n}G_{n}G_{n+k}$ is the function
obtained by convolving two exact Green functions with no vertex corrections. Thus  
\begin{equation}
\Lambda^{2}=\frac{\partial \chi _{k}/\partial (i\omega _{k})|_{k=0}}{%
\partial \chi _{k}^{0}/\partial (i\omega _{k})|_{k=0}}.  \label{l0fac}
\end{equation}
The right hand side of Eq. (\ref{l0fac}) can be calculated numerically by
QMC in the time domain because the leading long-time behavior comes from the first frequency derivative and is $\chi (0 \ll \tau \ll \beta) = (\pi/\beta) \partial \chi _{k}/\partial (i\omega _{k})|_{k=0}/\sin^2(\pi \tau /\beta )$. For $\beta \gtrsim 4$ we find $\chi(\tau)/\chi^0(\tau)$ is very flat in the center region of the interval $(0,\beta )$ (see inset to Fig. \ref{lam0u}).
The rapid decrease of $\Lambda$ again shows that the electron-phonon
mass enhancement and scattering rate are `turned off' near half filling and for strong correlations,
in disagreement  with the authors of Ref. \cite{Kim89} who found
a weaker decrease in the matrix element, which was moreover
cancelled by an increase in effective mass.

We summarize the main results of this paper. We have studied a general
electron-phonon model in a new approach that combines the adiabatic
expansion in $\gamma \ll 1$ of the conventional Migdal-Eliashberg
theory with a dynamical mean-field treatment of electron
correlations. Our main finding is an effective phonon action 
correct to ${\cal O}(\gamma^{3/2})$ with static
coefficients $\Gamma_N$ ($N = 2,3,4$) and vertex $\Lambda$, 
which are easy to compute numerically for various models. 
For the Holstein-Hubbard model we have shown that 
the electron-phonon coupling is  strongly suppressed
by Hubbard-$U$ effects in systems near Mott transitions.
Extension of our results to other correlated systems such as
heavy fermions would be worthwhile.

We thank S. Blawid for useful discussions, H. Monien and J. Freericks for critical reading of the manuscript, and NSF DMR00081075 and the University of Maryland/Rutgers MRSEC for support.

\end{multicols}
\end{document}